\documentstyle[prd,aps,floats]{revtex}
\begin{document}
\draft
%%%%%%%%%%%%%%%%%%%%%%%%%%%%%%%%%%%%%%%%%%%%%%%%%%%%%%%%%%%%%%%%%
%
%  Uncomment following four lines and one below for 2 column format
%  and figureinsertions.
%
\input epsf
% ELENA newcommands:
\newcommand{\Piso}{P_{\rm iso}}
\newcommand{\Pad}{P_{\rm ad}}
\newcommand{\Niso}{N_{\rm iso}}
\newcommand{\Nad}{N_{\rm ad}}
\newcommand{\Nmix}{N_{\rm mix}}
\newcommand{\Ndof}{N_{\rm dof}}
\newcommand{\niso}{n_{\rm iso}}
\newcommand{\nad}{n_{\rm ad}}
\newcommand{\khor}{k_{\rm hor}}
\newcommand{\kgal}{k_{\rm gal}}
\newcommand{\Tiso}{T_{\rm iso}}
\newcommand{\Tad}{T_{\rm ad}}
\newcommand{\Xmin}{\chi_{\rm min}}
\newcommand{\tfwhm}{\theta_{_{\rm fwhm}}}
\newcommand{\al}{\alpha}
\newcommand{\Bp}{B^{\prime}}
\newcommand{\HO}{H_0}
\newcommand{\omO}{\Omega_m}
\newcommand{\omL}{\Omega_\Lambda}
\newcommand{\omB}{\Omega_{_{\rm B}}}
\newcommand{\hm}{\,h^{-1}{\rm Mpc}}
\newcommand{\be}{\begin{equation}}
\newcommand{\ee}{\end{equation}}
\newcommand{\bea}{\begin{eqnarray}}
\newcommand{\eea}{\end{eqnarray}}

\renewcommand{\topfraction}{0.8}
\twocolumn[\hsize\textwidth\columnwidth\hsize\csname
@twocolumnfalse\endcsname
%%%%%%%%%%%%%%%%%%%%%%%%%%%%%%%%%%%%%%%%%%%%%%%%%%%%%%%%%%%%%%%%%%%%%%
\preprint{IMPERIAL-TP-98/99-72, UBC-COS-99-04, hep-ph/9909420}
\title{Microwave background anisotropies and large scale structure 
constraints \\ on isocurvature modes in a two-field model of inflation}
\author{Elena Pierpaoli}
\address{Department of Physics and Astronomy, University of
British Columbia, Vancouver, B.C. V6T 1Z1, Canada}
\author{Juan Garc{\'\i}a-Bellido}
\address{Theoretical Physics, Blackett Laboratory, Imperial College,
Prince Consort Road, London SW7 2BZ, U.K.}
\author{Stefano Borgani}
\address{INFN Sezione di Trieste, c/o Dipartimento di Astronomia
dell'Universit\`a, via Tiepolo 11, I-34131 Trieste, Italy\\
INFN Sezione di Perugia, c/o Dipartimento di Fisica 
dell'Universit\`a, Via A. Pascoli, I-06123 Perugia, Italy}

\date{November 17, 1999}
\maketitle

\begin{abstract}

In this paper we study the isocurvature mode contribution to the cosmic
microwave background anisotropies and the large scale structure power
spectrum, for a two-field model of inflation proposed by Linde and
Mukhanov. We provide constraints on the parameters of the model
by comparing its predictions with observations of the microwave
background anisotropies, large scale structure data on the galaxy
power spectrum, and the number density of nearby galaxy clusters. We
find that such models are consistent with observations for a narrow
range of parameters. As our main result, we find that only a very small
isocurvature component is allowed, $\al \le 0.006$, for any underlying
Friedmann model. Furthermore, we give the expected resolution with which
the model parameters will be determined from future satellite missions
like MAP and Planck, for a fiducial flat $\Lambda$CDM model. We find
that Planck mission will be able to detect such small contributions,
especially if polarization information is included. The isocurvature
spectral index $\niso$ will also be determined with better than $8\%$
precision.

\end{abstract}

\pacs{PACS number: 98.80.Cq \ %\hspace{1cm}
Preprint \ IMPERIAL-TP-98/99-72, UBC-COS-99-04, hep-ph/9909420}

%%%%%%% Comment the next line before submission
\vskip2pc]

\renewcommand{\thefootnote}{\arabic{footnote}}
\setcounter{footnote}{0}

\section{Introduction}

The future satellite experiments MAP \cite{MAP} and Planck \cite{Planck}
open an exciting era for cosmologists and for particle physicists. The
high resolution and sensitivity of these experiments will allow such a
precise determination of the cosmic microwave background (CMB) power
spectrum of temperature and polarization anisotropies that it will soon
be conceivable to test different cosmological models with great accuracy
\cite{paramest}.  Meanwhile, galaxy surveys, like 2dF and Sloan Digital
Sky Survey (SDSS), aimed at measuring several hundred of thousand
redshifts will provide a map of the Universe with unprecedented
precision and extension.  Any viable cosmological model must produce
reasonable fits to both the observed CMB and large scale structure (LSS)
power spectra, and must be tested on the basis of all available data.

Within the context of inflationary scenarios, Gaussian adiabatic
fluctuations are often assumed as a standard prediction. However,
besides the usual adiabatic fluctuations, other independent modes may be
present, e.g. isocurvature fluctuations \cite{LK86,EB86,KoSa87,HuSu95},
and in many cases they are non-Gaussian \cite{nongauss}. In particular,
the isocurvature mode is an entropy perturbation, characterized by an
appropriate balance of the fluctuations in the different components,
such that the spatial curvature remains unperturbed. There is an
attractive model that has been proposed some time ago that considers
pure isocurvature perturbations in the baryon component~\cite{Peebles},
but unfortunately seems to be ruled out by present
observations~\cite{HuBuSu95,FriGaz99}. Most of the recent models are
actually cold dark matter (CDM) isocurvature models~\cite{cdmiso}, but
there is also an ingenious neutrino isocurvature model~\cite{BuMoTu99}.

Far from being academic, the reason for considering also isocurvature
fluctuations resides in the fact that many different inflationary models,
with more than one scalar field, predict the formation of significant
isocurvature fluctuations during the inflationary era
\cite{LK86,modinfiso}. As for their predictions on the density perturbation
power spectra, they may differ among themselves in having different
amplitudes and tilts of isocurvature and adiabatic spectra, as well as
for the statistical nature of both modes.  Many inflationary models
predict that both the isocurvature and adiabatic fluctuations have a
nearly scale invariant (Harrison--Zeldovich) spectrum with Gaussian
statistics.  These models have been tested against LSS in the recent
literature \cite{stompor,KaSuYa98}, which showed that only a small
fraction of isocurvature component seems to be allowed by present
observations.

In this paper we focus on a particular inflationary scenario proposed by
Linde and Mukhanov \cite{LM97}, in which none of the two conditions
mentioned above is necessarily satisfied. In fact, here the isocurvature
fluctuations are non--Gaussian, more specifically, they are $\chi^2$
distributed,\footnote{However, slight variants of the model can also
give a Gaussian isocurvature mode~\cite{LM97}.} and their spectrum
is a power law, with spectral index $\niso>1$.  We test this model with
the present observations on CMB and LSS, and we also make predictions on
how well the future satellite experiments will be able to measure the
relevant model parameters.

In previous papers~\cite{stompor,KaSuYa98,EnKu99} the mixed spectra of
adiabatic plus isocurvature modes were assumed to be independent,
Gaussian and approximately scale invariant. Some~\cite{stompor,KaSuYa98}
used the large scale structura data, together with the COBE-DMR
normalization, to constrain the models, while others~\cite{EnKu99} only
used the CMB anisotropies, with or without polarization. In our paper,
we have combined both CMB and LSS observations, for a non scale
invariant spectrum of perturbations, and included also the 
gravitational wave contribution.

In section \ref{isofrominf} we review how isocurvature modes may arise
from inflation, and we describe the Linde--Mukhanov model which inspired
this work. In section \ref{powsp} we introduce the power spectra for
matter and radiation in the mixed (adiabatic + isocurvature) case,
showing the effect of the amplitude and tilt of the different
isocurvature contributions.  In section \ref{comwdata} we show the
comparison of the mixed spectra with the available data: we constrain
the parameter range considering a joint analysis of both CMB and LSS,
and we estimate with which precision the future satellite experiments
will determine the relevant parameters.  Finally, section
\ref{discussion} is dedicated to a general discussion of the results.

\section{Isocurvature modes from Inflation}\label{isofrominf}

Isocurvature perturbations are generated during inflation whenever there
is more than one scalar field present. They correspond to entropy
perturbations that do not perturb the metric, and thus the spatial
curvature. They typically arise when one of the fields is fixed by its
potential during inflation, the inflaton energy later decays into
relativistic particles and redshifts away, while the other field's
energy becomes the dominant contribution. Depending on model parameters,
the relative contribution of adiabatic to isocurvature perturbations may
be noticeable in the microwave background anisotropies and large scale
structure, and they may have in principle very different spectral tilts,
e.g. blue ($n>1$) or red ($n<1$). Furthermore, depending on the
evolution during inflation, the statistics of the different components
(adiabatic and isocurvature) could be very different, e.g. Gaussian and
$\chi^2$ distributed, respectively. Such a complicated phenomenology
requires a detail analysis in order to confirm whether a particular
model is ruled out by observations. 

In this paper we will concentrate in a particular realization of a
mixed adiabatic and isocurvature model proposed recently by Linde
and Mukhanov.

\subsection{The Linde--Mukhanov model}

The model of Ref.~\cite{LM97} is probably the simplest one can think of
that produces isocurvature perturbations during inflation.  It has two
coupled massive scalar fields described by the scalar potential, 
\be \label{vps} 
V(\phi,\sigma) = {1\over2} M^2\phi^2 + {1\over2} m^2\sigma^2 + 
{1\over2} g^2\sigma^2\phi^2 \,.  
\ee 
In principle, inflation could occur along either of the valleys at
$\phi=0$ or $\sigma=0$, depending on initial conditions.  In a chaotic
inflation approach one expects that the fields will start at very large
values, $\phi, \sigma\gg M_P$, where the coupling term
$g^2\sigma^2\phi^2$ dominates. Let us suppose that initially one of the
fields has a larger value, say $|\phi|>|\sigma|$ and thus the field
$\sigma$ rapidly settles at $\sigma=0$. Then inflation occurs along the
$\sigma=0$ valley, with energy density ${1\over2} M^2\phi^2$ and a
Hubble constant \be H^2 = {4\pi M^2\over3M_P^2}\,\phi^2\,.  \ee During
inflation the mass of the $\sigma$ field becomes \be \bar m^2 = m^2 +
\nu\,H^2 \ee where $\nu=3g^2M_P^2/4\pi M^2$ is a constant. Typically,
during inflation the second term dominates and thus the model gives a
mass term of the $\sigma$ field proportional to the rate of expansion.

The quasi-de-Sitter evolution during inflation provides a neat way to
generate metric perturbations from quantum fluctuations. Those of the
inflaton will give rise to adiabatic density perturbations, since the
energy density during inflation is proportional to the inflaton field
fluctuations, $\delta\rho\sim V'(\phi)\,\delta\phi$. On the other hand,
quantum fluctuations of the $\sigma$ field will not generate curvature
perturbations since the inflationary trajectory lies along
$\langle\sigma\rangle=0$.  Nevertheless, after inflation the energy
density of the $\sigma$ field may come to dominate the evolution of the
universe (e.g. as a cold dark matter component) and its fluctuations
would then contribute as isocurvature perturbations~\cite{LK86}. Let
us compute the amplitude of those perturbations. For a massive field
with $\bar m^2\ll H^2$ during inflation, the amplitude of the long
wavelength perturbation of the $\sigma$ field at the end of inflation is
given by
\be\label{sigk2}
k^3|\sigma_k^2| = {H^2\over2}\left({k\over H}\right)^{2\bar m^2/3H^2}\,,
\ee
and the average perturbations of energy density in the $\sigma$ field,
$\delta\rho_\sigma = \bar m^2(\delta\sigma)^2/2$ can be estimated 
as~\cite{LM97}
\be\label{drhosig}
{\delta\rho_\sigma\over\rho_{\rm tot}} \sim {\bar m^2\over M_P^2}\,
\left({k\over H}\right)^{2\bar m^2/3H^2}\,,
\ee
which corresponds to a ``blue'' spectral index 
\be
\niso \approx 1 + 4\bar m^2/3H^2 \simeq 1 + 4\nu/3>1\,.
\ee
The small ratio $\bar m^2/M_P^2$ ensures that the $\sigma$ field does not
contribute initially to the perturbations of the metric, 
i.e. it generates isocurvature perturbations, which much later
could end up dominating, as mentioned above. The detailed evolution is
very model dependent~\cite{LM97}. We will assume, following Linde
and Mukhanov, that the correction $\nu H^2$ to the mass of the $\sigma$
field disappears soon after inflation, when the inflaton field $\phi$
decays into relativistic particles while the $\sigma$ field remains
stable or decays very late, and thus its energy density (in coherent
oscillations of the field) may dominate in the form of cold dark matter
today. Under these assumptions one can estimate the corresponding
density contrast~\cite{LM97}
\be\label{densiso} 
{\delta\rho_\sigma\over\rho_\sigma} \sim C(k)\,
\left({k\over M}\right)^{2\nu/3}\,, 
\ee
for $k\ll {\rm min}[M\sqrt{m/M},\ M \exp(-1/2\nu)]$, where 
\be
C(k) = {\sqrt\nu\over\ln(M/m)+3/\nu}\,
\left[\ln\Big({M\over k}\Big)\right]^{1-\nu/3}\!
\left({M\over m}\right)^{2\nu/3}\,.
\ee
This isocurvature perturbation has non-Gaussian statistics, in fact a
$\chi^2$ distribution, because it arises from the square of a Gaussian
field [see Eq.~(\ref{drhosig})].
\footnote{One could avoid having non-Gaussian statistics for the
isocurvature component by giving the $\sigma$ field a vacuum expectation
value, $\langle\sigma\rangle\neq 0$. In this way, the main isocurvature
contribution would come from the term $\delta\rho_\sigma = \bar m^2
\langle\sigma\rangle \delta\sigma$, which is Gaussian distributed.}

On the other hand, the adiabatic density perturbations generated by the
inflaton field $\phi$ during inflation contribute with a Gaussian
spectrum with amplitude
\be\label{densad}
{\delta\rho_k\over\rho} \simeq \sqrt{4\over3\pi}\,N_k\,{M\over M_P}\,
\left({k\over aH}\right)^{-2/N_k}\,,
\ee
where $N_k= N_{\rm hor} = 65$ is the number of e-folds before the
end of inflation when the mode with wavenumber $k$ corresponding to
our present horizon crossed the Hubble scale during inflation,
with the spectral index
\be\label{nad}
\nad = 1 - {2\over N_k}\,. 
\ee

There is enough freedom in the model parameters $(\nu, M, m)$ to have
the Gaussian adiabatic perturbations dominate the spectrum on large
scales, while the non-Gaussian isocurvature perturbations dominate on
small scales.  In practice, the latter will never dominate at scales of
cosmological interest since, as we shall illustrate in Sections IV and
V, the relative contribution of the isocurvature component is
constrained by present observations to be very small.

We will thus consider a mixed model of adiabatic and isocurvature
perturbations arising from inflation and contributing simultaneously
to the temperature anisotropies of the microwave background and
the power spectrum of large scale structure. The relative tilts
and amplitudes of the two spectra will allow us to compare with
present observations and determine the likelihood functions for
a given set of model parameters.

We will use the CMBFAST code~\cite{SelZal96,cmbfast} to normalize the
model to COBE data \cite{COBE,Bond,BuWh97}, compute the $C_l$ and power
spectra $P(k)$, and then compare with observations.

Therefore, the post--inflationary power spectra for the independent
adiabatic and isocurvature components read 
\bea 
\Pad(k) =
25\times10^{-10}\,\left({k\over\khor}\right)^{\nad}\,,\\ 
\Piso(k)=
d_{10}\times10^{-10}\,\left({k\over\kgal}\right)^{\niso}\,, 
\eea 
where $d_{10}= {\cal O}(1 - 20)$ is an arbitrary normalization factor,
to be fixed by observations. We will discuss in the next Section the
relative normalization of the two components for a mixed power spectrum.

\section{Matter and Radiation Power Spectra}\label{powsp}

In order to compare the theory with the data, we compute the power
spectra of matter and radiation at the present epoch. For this purpose,
we solve Boltzmann equations using CMBFAST, starting from a very early
epoch ($z \simeq 10^7$) up to the present time.

In the presence of only one mode of fluctuations the actual matter power
spectrum is: 
\be 
P(k)\,=\,T^2(k)\,P_{\rm in}(k)\,, 
\ee 
where $P_{\rm in}(k)$ denotes the initial power spectrum and the
transfer function $T(k)$ contains the information on how the spectrum is
modified through the evolution. $T(k)$ is typically of order 1 on large
scales (small $k$) and it becomes less than 1 at smaller scales (larger
$k$), the actual shape depending on the specific mode considered (i.e.,
whether adiabatic or isocurvature), the cosmological parameters and
on the nature of the dark matter content. In our analysis, while we will
allow for both adiabatic and isocurvature fluctuations for general
Friedmann background, we will assume the dark matter content to be
contributed only by CDM and baryons, ignoring any contribution that may 
come from neutrinos in the form of hot dark matter.

%%%%%%%%%%%%%%%%%%%%%%%%%%%%%%%%%%%%%%%%%%%%%%%%%%%%%%%%%%%%%%%%%%%%
%   FIGURE 1 : T(K)
%%%%%%%%%%%%%%%%%%%%%%%%%%%%%%%%%%%%%%%%%%%%%%%%%%%%%%%%%%%%%%%%%%%
\begin{figure}[t]
\centering
\hspace*{-4.5mm}
\leavevmode\epsfysize=8.5cm \epsfbox{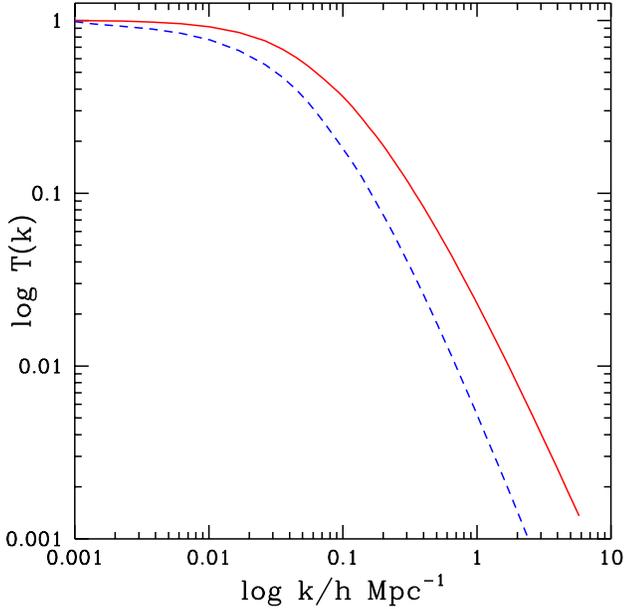}\\[3mm]
\caption[fig1]{\label{fig1} Transfer function for the adiabatic (solid
curve) and isocurvature (dashed curve) modes in the standard cold dark
matter model ($\omO =1$, $h = 0.5$ and $\omB = 0.05$). The
isocurvature mode is more damped on small scales. }
\end{figure}
%%%%%%%%%%%%%%%%%%%%%%%%%%%%%%%%%%%%%%%%%%%%%%%%%%%%%%%%%%%%%5%%%%%%

%%%%%%%%%%%%%%%%%%%%%%%%%%%%%%%%%%%%%%%%%%%%%%%%%%%%%%%%%%%%%%%%%%%%%
%  FIGURE  2:  P(K)
%%%%%%%%%%%%%%%%%%%%%%%%%%%%%%%%%%%%%%%%%%%%%%%%%%%%%%%%%%%%%%%%%%%%%
\begin{figure}[t]
\centering
\hspace*{-6mm}
\leavevmode\epsfysize=9cm \epsfbox{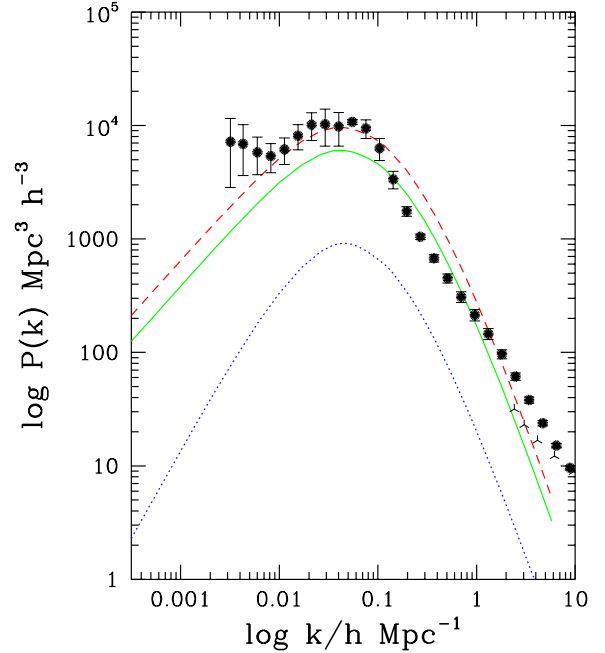}\\[3mm]
\caption[fig2]{\label{fig2} Power spectra of pure isocurvature (dotted
curve) and adiabatic (dashed curve) modes with spectral indices $\nad
= 0.962$ and $\niso = 1.56$. The middle (solid curve) power spectrum
is a mixed one, corresponding to $\al = 8 \times 10^{-3}$. The data
points are from the APM survey.}
\end{figure}
%%%%%%%%%%%%%%%%%%%%%%%%%%%%%%%%%%%%%%%%%%%%%%%%%%%%%%%%%%%%%%%%%%%%

As for the adiabatic and isocurvature transfer functions, we take the
CDM expressions
\bea
&&\Tad(k) = {\ln(1+2.34q)\over
2.34q}\,\times \nonumber \\ 
&&\Big[1+3.89q+(16.1q)^2+(5.46q)^3+(6.71q)^4 \Big]^{-1/4}\,;
\label{eq:tad} \\
&&\Tiso(k) = \nonumber \\
&&\Big[1+{(40q)^2\over
1+215q+(16q)^2(1+0.5q)^{-1}}+(5.6q)^{8/5}\Big]^{-5/4} \,,
\label{eq:tiso}
\eea 
as provided by Bardeen et al. \cite{bbks}, where $q=k/\Gamma\,$
$(\hm)^{-1}$. Here $\Gamma$ is the so called ``shape'' parameter and for
CDM model is defined as $\Gamma=h\omO \exp(-\omB-\sqrt{2h}\omB/\omO)$,
so as to account for the small--scale damping due to the presence of a
non--negligible baryon fraction \cite{sugi95}. Note that we have
redefined Eq. (\ref{eq:tiso}) with respect to the expression given in
Ref. \cite{bbks} so as to have the same small--$k$ asymptotic behavior,
$T(k)\to 1$, for both components. Among other things, this means that
$\niso=1$ corresponds to a scale invariant spectrum, instead of the
usual $\niso=-3$ appearing in the literature.  We verified that
Eqs. (\ref{eq:tad}) and (\ref{eq:tiso}) reproduce quite well the outputs
of the CMBFAST code for the $k$--range of interest for our analysis.

As an example, we show in Figure \ref{fig1} the shape of these two
transfer functions for a particular choice of the cosmological
parameters. It is apparent that the isocurvature case has a larger
damping at large $k$ values, due to the fact that they start out as
zero-curvature or isothermal fluctuations and takes longer than the
adiabatic ones to build up, after the matter--radiation equality epoch.

Since adiabatic and isocurvature are independent modes of fluctuations,
when both modes are present at the same time the total power spectra of
matter and radiation can be computed as a linear combination of the pure
isocurvature and adiabatic ones.  In the LM model considered in this
work, both the adiabatic and isocurvature modes have power law initial
power spectra, but with different spectral indices. Therefore, the overall
spectrum can be casted in the form
\be \label{Pmix}
P_{\rm mix}(k) = \Nmix \left[
(k\tau_0)^{\nad} \Tad^2(k) + 
\al\,(k\tau_0)^{\niso} \Tiso^2(k) \right] \,,
\ee
where $\tau_0$ is the conformal time at present and $\al$ is a
dimensionless parameter that indicates the relative contribution of
isocurvature and adiabatic perturbations (pure adiabatic and
isocurvature power spectra correspond to $\al = 0$ and $\al = \infty$
respectively).  The comparison of our $\alpha$ with the mixing
coefficients introduced by other authors
(e.g. \cite{stompor,KaSuYa98,EnKu99}) may not be straightforward
because, in contrast with their approach, we have also allowed for $\nad
\ne \niso \ne 1$.  In the case $\nad = \niso = 1$, our $\al$ compares
with the other choices as follows: $\al = (1 - \al_S)/\al_S$
\cite{stompor}, $\al = \al_{KKSY}$ \cite{KaSuYa98}, $\al=\al_{EK}/(1 -
\al_{EK})$ \cite{EnKu99}.  In eq.(\ref{Pmix}), $\Nmix$ is the
normalization factor, that we derived by normalizing the CMB spectrum to
the COBE data \cite{Bond,BuWh97}.  Different mixing coefficients $\al$ lead
to different normalizations:
\be 
\Nmix = {\Nad\over {1 + \al f}}\,.  
\ee
In the above expression, $f={\Nad/\Niso}$ where $\Nad$ and $\Niso$ are
the normalizations of the pure adiabatic and isocurvature scalar modes, and
their ratio conveys the information about the different Sachs--Wolfe
contribution of isocurvature and adiabatic modes to CMB anisotropies. In
the normalization, we also took into account the tensor contribution to
the anisotropies, according to the parameter specified below.  A typical
hybrid spectrum is plotted in fig.~\ref{fig2}, together with the APM
data points \cite{pow2d}.  Given the big tilt of the isocurvature power
spectrum, the mixed one has a slope similar to the adiabatic one at
small $k$.  The effect of the different normalization is evident even
for low values of the $\al$ coefficient.

%%%%%%%%%%%%%%%%%%%%%%%%%%%%%%%%%%%%%%%%%%%%%%%%%%%%%%%%%%%%%%%%%%%%%%%%%
%    figure 3: mixed C_l as a function of \alpha (fixed n_iso)
%%%%%%%%%%%%%%%%%%%%%%%%%%%%%%%%%%%%%%%%%%%%%%%%%%%%%%%%%%%%%%%%%%%%%%%
\begin{figure}[t]
\centering
\hspace*{-9mm}
\leavevmode\epsfysize=8.5cm \epsfbox{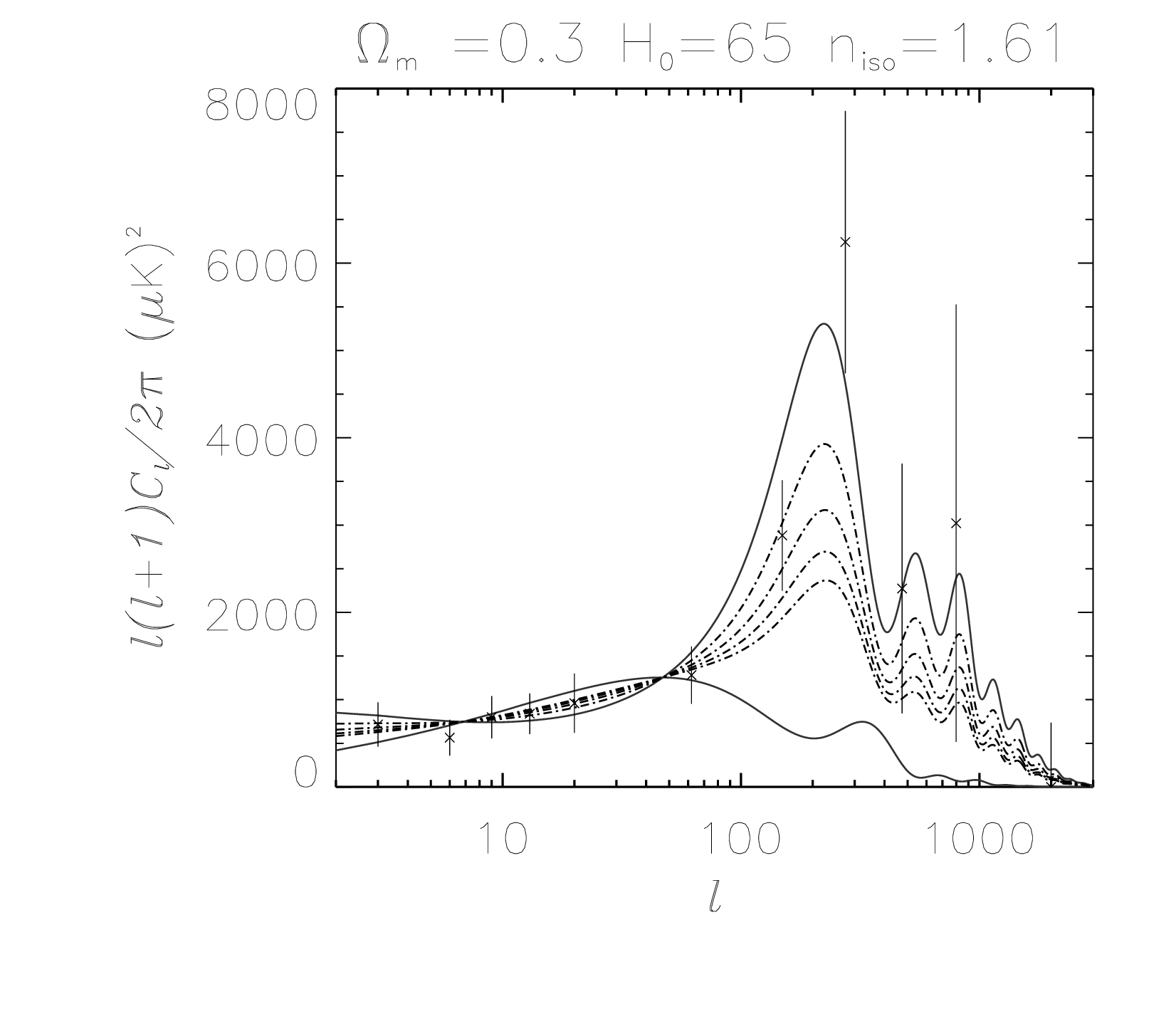}\\[3mm]
\caption[fig3]{\label{fig3} Mixed CMB spectra with fixed $\niso$ value.
The solid lines correspond to the pure adiabatic and isocurvature modes,
normalized to COBE. The dotted--dashed lines corresponds to $\al =
3, 6, 9, 12 \times 10^{-3}$ from top to bottom. The points shown here
are the binned spectrum from Ref.~\cite{Bond98}. However, we have used 
the whole set of experiments from Table \ref{tab:esp} in the CMB analysis.} 
\end{figure}

%%%%%%%%%%%%%%%%%%%%%%%%%%%%%%%%%%%%%%%%%%%%%%%%%%%%%%%%%%%%%%%%%%%%%%

%%%%%%%%%%%%%%%%%%%%%%%%%%%%%%%%%%%%%%%%%%%%%%%%%%%%%%%%%%%%%%%%%%%%%%%%%
%    figure 4: mixed C_l as a function of \n_iso (fixed \alpha)
%%%%%%%%%%%%%%%%%%%%%%%%%%%%%%%%%%%%%%%%%%%%%%%%%%%%%%%%%%%%%%%%%%%%%%%
\begin{figure}[t]
\centering
\hspace*{-6mm}
\leavevmode\epsfysize=8.5cm \epsfbox{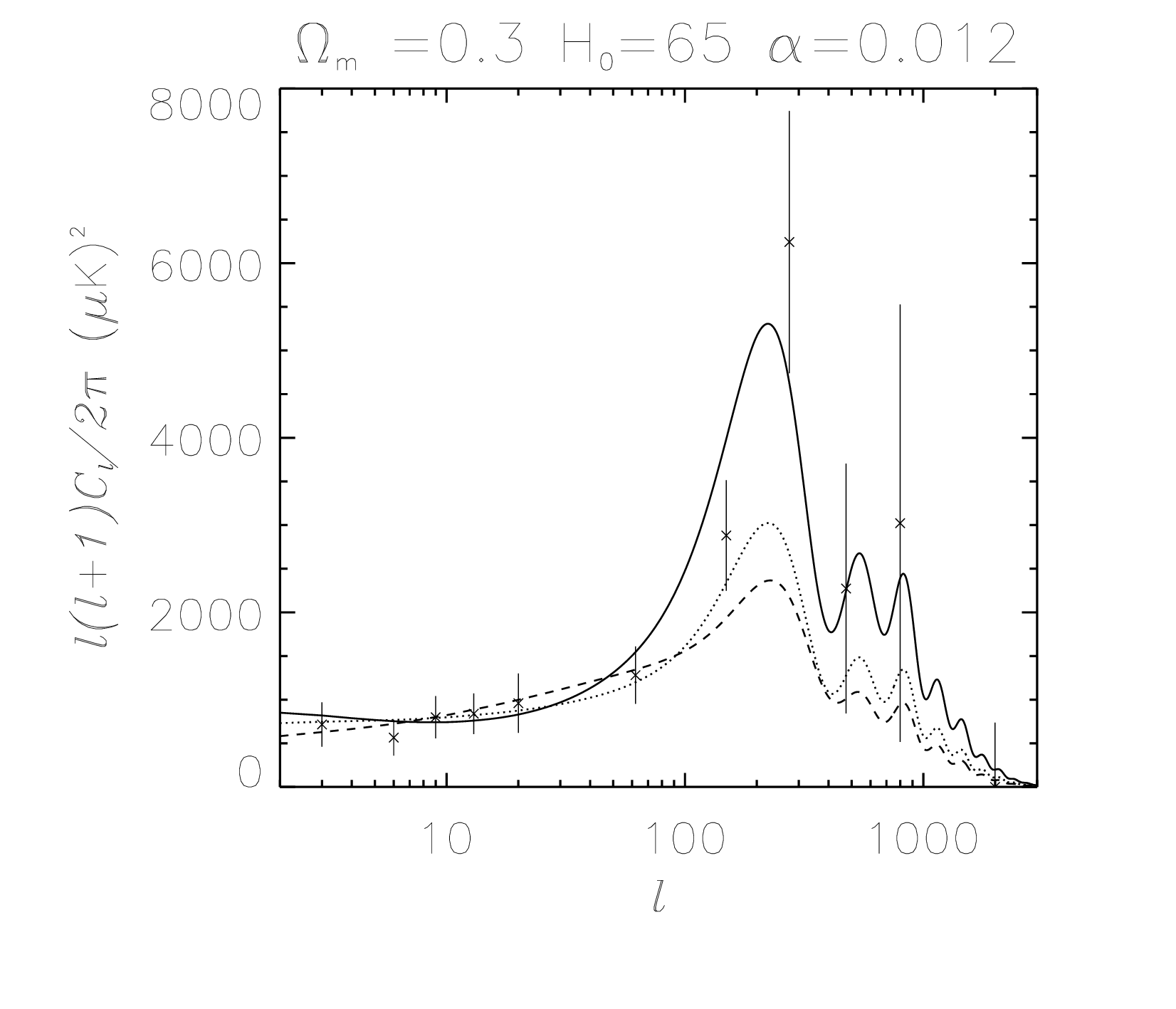}\\[3mm]
\caption[fig3]{\label{fig4} Mixed $C_l$ spectra with different tilts of
the isocurvature component with fixed ratio $\al=0.012$. The solid line
is the pure adiabatic mode ($\nad = 0.962$), the dotted line corresponds
to mixed isocurvature ($\al=0.012$) with $\niso = 1.36$ and the dashed
line to $\niso = 1.61$.}
\end{figure}
%%%%%%%%%%%%%%%%%%%%%%%%%%%%%%%%%%%%%%%%%%%%%%%%%%%%%%%%%%%%%%%%%%%%%%%%

As for the CMB anisotropies, following the standard notation, we
describe their dependence on the direction $\hat{n}$ as
\be
{\Delta T(\hat{n}) \over T_o} = \sum_{l=0}^{\infty}\, \sum_{m=-l}^{l}
a_{lm} \ Y_{lm}(\hat{n})\,.
\ee
Therefore, the radiation power spectrum is defined as
\be
C_l=\langle| a_{lm} |^2 \rangle\,,
\label{eq:cldef}
\ee 
where the brackets denote the ensemble average over different
realizations. When both scalar and tensor modes are present, 
the $C_l$ can be decomposed as $C_l = C_l^S + C_l^T$. 
In Linde--Mukhanov model the tensor contribution is included in the
adiabatic mode.

 Similarly to what happens to the matter power spectrum,
the radiation power spectrum for the mixed case can be found as a linear
combination of the two independent adiabatic and isocurvature power spectra
$C_l^{\rm ad}$ and $C_l^{\rm iso}$ : 
\be
C_l^{\rm mix}= {{C_l^{\rm ad} + f \al C_l^{\rm iso}} \over { 1 + f \al}} \,.
\ee 
In the expression above $C_l^{\rm ad}$ and $C_l^{\rm iso}$ are both
normalized to COBE, according to ref.\cite{BuWh97}. In computing the
power spectra, we modified the CMBFAST code \cite{SelZal96,cmbfast} to
our purposes.  In figures \ref{fig3} and \ref{fig4} we show some
examples of CMB spectra, together with the estimated binned spectrum
from Bond et al. \cite{Bond98}, which provides a visual indication of
the experimental status. However, we have used the whole set of
experiments from Table \ref{tab:esp} in the CMB analysis. The effect of
adding an isocurvature component is to add a lot of power on small
multipoles through the SW effect. The anti--tilt of the spectrum is in
general not enough to compensate this effect, and the first acoustic
peak is consequently lower than in the corresponding pure adiabatic
case. Fig.~\ref{fig3} shows examples of mixed spectra with different
$\al$ values for fixed $\niso$ and $\nad$. It shows that even a small
$\al$ can cause a significant damping of the acoustic peaks.  Moreover
note that a high $\niso$ component also leads to suppressed acoustic
peaks, for a fixed $\al$ value (see Fig.~\ref{fig4}).  This is because
the bigger is $\niso$ the smaller is the normalization coefficient
$\Niso$ and the bigger is $f$.  Therefore for fixed $\al$, the
isocurvature $C_l$ spectrum takes more weight in the mixture.

\subsection{Choice of model parameters}

\begin{figure}[t]
\centering
\hspace*{-5.5mm}
\leavevmode\epsfysize=6.5cm \epsfbox{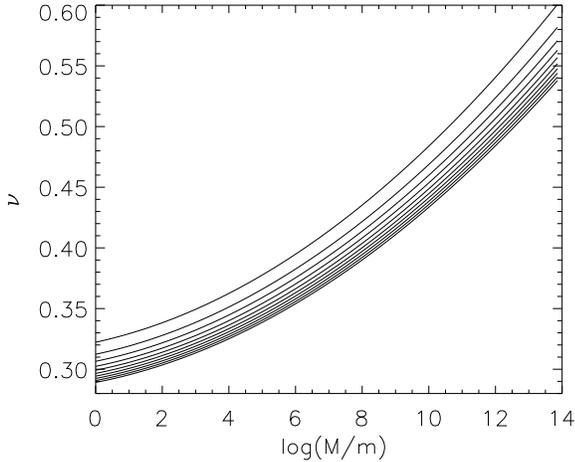}\\[3mm]
\caption[fig6]{\label{fig6} Typical values of $\nu$ as a function
of $M/m$, with $d_{10}$ in the range 1--10, from top to
bottom.}
\end{figure}

We will consider here a set of values for the parameters of the model.
Let us start with the scalar spectrum. The adiabatic tilt is given by
Eq. (\ref{nad}),
\be
\nad = 1 - 2/N_{\rm hor} = 0.9692 \,,
\ee
for $N_{\rm hor} = 65$, while the
spectral index for isocurvature fluctuations is
\be
\niso = 0.9722 + 4\nu/3 \,.
\ee
Figure~\ref{fig6} shows some typical values of the parameter
$\nu$ as a function of $M/m$, for different choices of $d_{10}$.

%%%%%%%%%%%%%%%%%%%%%%%%%%%%%%%%%%%%%%%%%%%%%%%%%%%%%%%%%%%%%%%%%%%%%%%
%    FIGURE 6 : NU(M/M)
%%%%%%%%%%%%%%%%%%%%%%%%%%%%%%%%%%%%%%%%%%%%%%%%%%%%%%%%%%%%%%%%%%%%%%%

%%%%%%%%%%%%%%%%%%%%%%%%%%%%%%%%%%%%%%%%%%%%%%%%%%%%%%%%%%%%%%%%%%%%%%

As for the tensor contribution, we considered:
\be
n_T = - 1/N_{\rm hor} = -0.0154 \,,
\ee
and a tensor to scalar ratio given by:
\be
R = {C_2^T \over C_2^S} \simeq -7n_T = 0.1077 \,,
\ee
which is not negligible.  In Fig. \ref{fig4} the tensor contribution is
very small and is already included in the solid line for the adiabatic
component, in order to make emphasis on its difference with respect to
the isocurvature component.

\begin{figure}[t]
\centering
\hspace*{-5.5mm}
\leavevmode\epsfysize=6.5cm \epsfbox{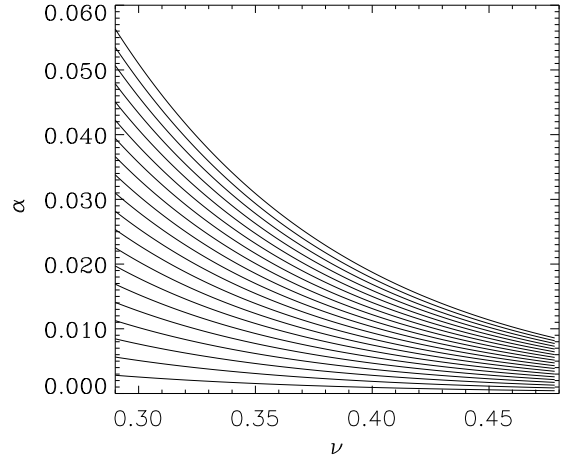}\\[3mm]
\caption[fig5]{\label{fig5} Typical values of $\al$ as a function
of the tilt parameter $\nu$ in the isocurvature spectrum. The different 
curves refer to different $d_{10}$ ranging from 1 to 20, from bottom to
top. }
\end{figure}

With the above spectral properties, we determined the relative amplitude
$\al$ in (\ref{Pmix}) to be related to the input parameters $d_{10}$
and $\nu$,
\be\label{ald10}
\al = d_{10} \times 10^{-1.291-4.343~\nu} \,.
\ee
For any value of $d_{10}$ between 1 and 10, the tilt parameter $\nu$
ranges between 0.29 and 0.48. As a consequence, $\niso$ is found to be
between 1.36 and 1.61.\footnote{Note that this is equivalent to $-2.64 <
\niso < -2.39$ in the usual notation, where $\niso=-3$ is the scale
invariant isocurvature perturbation.}  Note that while $\nu$ fixes the
value of the isocurvature spectral index, several $\al$ values are still
possible, depending on the $d_{10}$ values. In fig.~\ref{fig5} we plot
the value of $\al$ as a function of $\nu$, for $d_{10}$ in the range $1
- 20$. In any case, the values of $\al$ found are small, and for $d_{10}
< 32$, $\al$ never exceeds 0.09. In the comparison with the data, we
considered $\al < 0.08 $ and $1.36 < \niso < 1.61$, and treated them as
independent parameters, although they are not really independent [see
Eq. (\ref{ald10})]. 

%%%%%%%%%%%%%%%%%%%%%%%%%%%%%%%%%%%%%%%%%%%%%%%%%%%%%%%%%%%%%%%%%%%%%%
% FIGURE 5: ALPHA(NU)
%%%%%%%%%%%%%%%%%%%%%%%%%%%%%%%%%%%%%%%%%%%%%%%%%%%%%%%%%%%%%%%%%%%%
%%%%%%%%%%%%%%%%%%%%%%%%%%%%%%%%%%%%%%%%%%%%%%%%%%%%%%%%%%%%%%%%%%%%%%

\begin{figure}[t]
\centering
\hspace*{-1.8cm}
\leavevmode\epsfysize=7.2cm \epsfbox{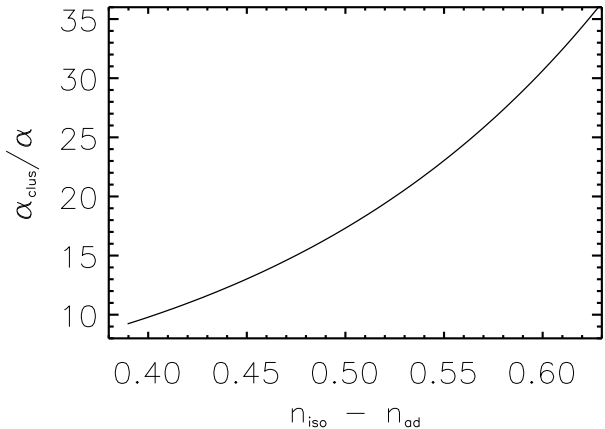}\\[-3mm]
\caption[fig9]{\label{fig9} The ratio $\alpha_{\rm clus}/\alpha$ 
between the relative isocurvature contribution $\alpha$ at the
scale of clusters and at COBE scales, as a function of the tilt
difference, $\niso-\nad$. }
\end{figure}

Note that we have represented in Fig.~\ref{fig5} the relative
contribution $\alpha$ of the isocurvature component to the total power
spectrum at COBE scales, i.e. {\em large scales}. Due to the strong
positive tilt of the isocurvature spectrum ($\niso>1$), this relative
contribution {\em increases} towards smaller scales. Since one of the
observational constraints that we will consider in the following is
represented by the number density of local galaxy clusters, it is
interesting to estimate the isocurvature contribution at the
characteristic cluster scales, $\sim 10\hm$. To this purpose we
introduce the quantity 
\be 
\alpha_{\rm clus}/\alpha = (k_{\rm clus}/k_{\rm hor})^{\niso-\nad}=
(300)^{\niso-\nad}\,,
\label{eq:alcl}
\ee
which provides the isocurvature contribution at the scale of galaxy
clusters. We plot the cluster-- to large--scale ratio of the
isocurvature fraction in Fig.~\ref{fig9}. It is apparent that the
enhancement of the isocurvature contribution can be rather large, such
that its effect is not negligible at the cluster scale, even for
a rather small large--scale contribution.

We will discuss in Section \ref{discussion} the 
implications of our results on the Linde--Mukhanov model.
As for the cosmological parameters $\omO$, $\omL$ and $\HO$, we always
considered flat models ($\omO +\omL = 1 $), with $ 0.2 \leq \omO \leq
1$, and $\HO \equiv 100\,h = 50, 65, 80$ km s$^{-1}$ Mpc$^{-1}$. As
for the density parameter contributed by baryons, we take the value
$\omB = 0.019\,h^{-2}$, which follows from the low deuterium abundance,
as determined by Burles \& Tytler \cite{BT98}.

%%%%%%%%%%%%%%%%%%%%%%%%%%%%%%%%%%%%%%%%%%%%%%%%%%%%%%%%%%%%%%%%%%%%%
% FIGURE 9: ALPHAFACTOR
%%%%%%%%%%%%%%%%%%%%%%%%%%%%%%%%%%%%%%%%%%%%%%%%%%%%%%%%%%%%%%%%%%%
%%%%%%%%%%%%%%%%%%%%%%%%%%%%%%%%%%%%%%%%%%%%%%%%%%%%%%%%%%%%%%%%%%%%%%

\section{Observational constraints}\label{comwdata}

In this section we test our model against the available microwave
background and large scale structure data, and we make predictions on
the precision with which future satellite experiments will determine the
relevant parameters. To this purpose, we use the Fisher matrix technique
to determine how well MAP and Planck satellites will constrain the
isocurvature contribution to the total power spectrum and its spectral
index.

\subsection{The microwave background data}

In the comparison of the CMB data with the model predictions, we
performed a $\chi^2$ analysis, first applied to CMB data by Lineweaver
et al. \cite{Lw97}. More precisely, we computed the $\chi^2$, as a
function of the cosmological parameters $\vec \lambda$, on the
band--power estimates of the CMB data, $\delta T^{\rm obs}(n)$, and the
model predictions, $\delta T^{\rm mod} (\vec \lambda, n)$, given $N_{\rm
exp}$ observed data points with their errors $\sigma$ :
\be 
\chi^2(\vec \lambda) = \sum_{n=1}^{N_{\rm exp}} \left[{\delta
T^{\rm obs}(n) - \delta T^{\rm mod}( \vec \lambda, n) } \over { \sigma (n) }
\right]^2.
\label{eq:chi2}
\ee
Evaluating the expression above, we used the 41 data points
\cite{CMBexplit} listed in table \ref{tab:esp}, and the corresponding
window functions.  We chose not to introduce the recent points from
MAT \cite{MATpap} and Python V \cite{Coble99} experiments because the
former does not yet include calibration errors and the results of the
latter are still under discussion.  
The value of $\chi^2$ obtained is
a function of the model parameters $\vec \lambda$.

%%%%%%  QUI VA LA TABELLA DEGLI ESPERIMENTI %%%%%%%%%%%%%%%%%%%%%
\begin{table}[h]
\centering
\begin{tabular}{cccc}
Experiment & $\Delta T_l \pm \sigma (\mu K)$ & $l_{\rm eff}$ & reference \\
\hline \\[-2mm]
COBE & $8.5^{+16}_{-8.5}$ &  2.1 & Tegmark \& Hamilton (1997)  \\      
COBE & $28.0^{+7.4}_{-10.4}$ & 3.1 &  Tegmark \& Hamilton (1997) \\
COBE  & $34.0^{+5.9}_{-7.2}$ & 4.1 & Tegmark \& Hamilton (1997)  \\
COBE & $25.1^{+5.2}_{-6.6}$ &  5.6 &  Tegmark \& Hamilton (1997)   \\
COBE & $29.4^{+3.6}_{-4.1}$ &  8.0 &   Tegmark \& Hamilton (1997)  \\
COBE & $27.7^{+3.9}_{-4.5}$ &  10.9 & Tegmark \& Hamilton (1997) \\
COBE & $26.1^{+4.4}_{5.3}$  &  14.3 & Tegmark \& Hamilton (1997) \\
COBE & $33.0^{+4.6}_{-5.4}$ &  19.4 & Tegmark \& Hamilton (1997) \\
SASK & $49.0^{+8.0}_{-5.0}$ &  86 & Netterfield et al (1997) \\
SASK & $69.0^{+7.0}_{-6.0}$ &  166 & Netterfield et al (1997) \\
SASK & $85.0^{+10.0}_{-8.0}$ &   236   & Netterfield et al (1997)\\   
SASK & $ 86.0^{+12.0}_{-10.0}$ & 285  & Netterfield et al (1997) \\   
SASK & $69.0^{+19.0}_{28.0}$ &  348  &  Netterfield et al (1997)\\   
CAT  & $50.8^{+15.4}_{-15.4}$ &  396 &   Scott et al (1996) \\
CAT  &$49.0^{+16.9}_{-16.9}$& 608 & Scott et al (1996) \\
CAT  &$57.3^{+10.9}_{-13.6}$& 415 &  Baker et al (1998) \\
FIRS &$29.4^{+7.8}_{-7.7}$&   10 & Ganga et al (1994) \\
TENERIFE &$34.1^{+12.5}_{-12.5}$& 20 & Hancock et al (1997) \\ 
SP91 &$30.2^{+8.9}_{-5.5}$&   57 &  Gundersen et al (1995)\\
SP94 &$36.3^{+13.6}_{-6.1}$&  57 & Gundersen et al (1995)\\
BAM &$48.4^{+16.5}_{-16.5}$& 74 & Tucker et al (1997) \\
ARGO &$39.1^{+8.7}_{-8.7}$& 95 & de Bernardis et al (1994) \\
ARGO &$46.8^{+9.5}_{-12.1}$ &95 & Masi et al (1996) \\
MAX-GUM &$54.5^{+16.4}_{-10.9}$& 145 &  Tanaka et al (1996)\\
MAX-ID &$46.3^{+21.8}_{-13.6}$& 145 &  Tanaka et al (1996) \\
MAX-SH &$49.1^{+21.8}_{-16.4}$& 145 & Tanaka et al (1996)  \\
MAX-HR &$32.7^{+10.9}_{-8.2}$& 145 &  Tanaka et al (1996)\\
MAX-PH &$51.8^{+19.1}_{-10.9}$& 145 &  Tanaka et al (1996) \\
PYTHON1 &$54.0^{+14.0}_{-12.0}$& 92 & Platt et al (1996) \\
PYTHON2 &$58.0^{+15.0}_{-13.0}$& 177 & Platt et al (1996)  \\
IAC &$112.0^{+65.0}_{-60.0}$&   33 & Femenia et al (1998) \\
IAC &$55.0^{+27.0}_{-22.0}$& 53 &  Femenia et al (1998) \\
IAB &$94.5^{+41.8}_{-41.8}$& 125 &  Piccirillo \& Calisse (93) \\ 
MSAM & $62.0^{+21.7}_{-21.7}$& 143 & Cheng et al (1996) \\
MSAM &$60.4^{+20.1}_{-20.1}$&  249 &  Cheng et al (1996)\\
MSAM &$50.0^{+16.0}_{-11.0}$&  160 & Cheng et al (1997) \\
MSAM &$65.0^{+18.0}_{-13.0}$& 270 & Cheng et al (1997) \\
QMAP &$47.0^{+6.}_{-7.}$&  80 & deOliveira--Costa et al (98) \\
QMAP &$59.0^{+6.}_{-7.}$&  126 &  deOliveira--Costa et al (98) \\
QMAP &$52.0^{+5.}_{-5.}$&  111 & deOliveira--Costa et al (98)  \\
OVRO &$56.0^{+8.5}_{-6.6}$& 589 & Leitch et al (1998) \\
\end{tabular}
\vspace*{2mm}
\caption{Data points used in the $\chi^2$ analysis. First column is the
experiment, second column is the experimental value with its error,
third column is the effective multipole number, and fourth column is the
reference paper.  }
\label{tab:esp}
\end{table}  
%%%%%%%%%%%%%%%%%%%%%%%%%%%%%%%%%%%%%%%%%%%%%%%%%%%%%%%%%%%%%%%%%%

For each choice of the parameters $\Omega_m$ and $\HO$, which describe
the Friedmann background, we compute the $\chi^2$ between model and data
in the $(\niso,\al)$ parameter space.  In Table \ref{tab:chimin} we
report the values of the parameters of the best fit, the corresponding
$\Xmin^2$ value, and the probability of getting that $\chi^2$ value with
present data if the model considered is the real one.

For $h=0.5$ a small contribution of the isocurvature component is
desirable, especially for low $\omO$ universes.  On the other hand, for
$h=0.8$ the addition of an isocurvature contribution to the adiabatic
mode doesn't provide a better fit to the data.  In general, however, the
allowed isocurvature fraction tends to be small ($\al<0.01$), while the
isocurvature spectral index is not significantly
constrained (see Fig.~\ref{fig4}). In any case, the best fit to the
data is provided by the lowest $\niso$ considered. 
  
%%%%%%%%%%%%%%%%%%%%%%%%%%%%%%%%%%%%%%%%%%%%%%%%%%%%%%%%%%%%%%%%%%%%%%%%%
%			TABLE OF CHI^2 VALUES
%%%%%%%%%%%%%%%%%%%%%%%%%%%%%%%%%%%%%%%%%%%%%%%%%%%%%%%%%%%%%%%%%%%%%%%%
\begin{table}[h]
\centering
\begin{tabular}{ccccc}
$\omO$ & $\niso$ & $\al$ &
$\Xmin$ & $P(\chi < \chi_{\rm min})$ \\ 
\hline\\[-2mm]
0.2  &    1.359  & 0.006    &    21.6  &    0.015 \\
0.3  &    1.359  & 0.003    &    21.9  &    0.017 \\
0.4  &    1.359  & 0.001    &    22.7  &    0.023 \\
0.5  &    1.359  & 0        &    23.5  &    0.032 \\
0.6  &    1.359  & 0        &    24.8  &    0.048 \\
0.7  &    1.359  & 0        &    26.4  &    0.078 \\
0.8  &    1.359  & 0        &    28.3  &    0.126 \\
0.9  &    1.359  & 0        &    30.5  &    0.200 \\
1    &    1.359  & 0        &    33.5  &    0.323 \\
\hline
%\hline
0.2 &      1.359 & 0.001   &     24.4   &   0.043 \\
0.3 &      1.359 & 0       &     26.4   &   0.08   \\
0.4 &      1.359 &     0   &     29.2   &   0.15 \\
0.5 &      1.359 &     0    &    32.7   &   0.29 \\
0.6 &      1.359 &     0    &    37.4   &   0.50 \\
0.7 &      1.359 &     0    &    43.1   &   0.74 \\
0.8 &      1.359 &     0    &    49.2   &   0.89 \\
0.9 &      1.359 &     0    &    55.5   &  0.97 \\
1   &      1.359 &     0    &    61.8   &  0.99 \\
\hline
%\hline

0.2 &      1.359   &   0   &   30.2  &   0.19 \\
0.3 &      1.359   &   0   &   36.0  &   0.44 \\
0.4 &      1.359   &   0   &   44.1  &   0.77 \\
0.5 &      1.359   &   0   &   52.8  &   0.94 \\
0.6 &      1.359   &   0   &   61.5  &   0.99 \\
0.7 &      1.359   &   0   &   69.8  &   0.99 \\
0.8 &      1.359   &   0   &   77.8  &   0.99 \\
0.9 &      1.359   &   0   &   85.4  &   0.99 \\
1   &      1.359   &   0   &  92.6   &   0.99 \\
\end{tabular}
\vspace*{2mm}
\caption{Results of the $\chi^2$ analysis for different models
($h=0.50, 0.65, 0.80$ from top to bottom). Column 1: the matter density
parameter. Column 2 and 3: the best fit values for $\niso$ and $\al$
respectively. Column 4: the corresponding $\Xmin^2$ value. Column 5: the
probability of getting a smaller $\chi^2$, assuming that this is the
real model.  We considered here 41 experimental points and 3 free
parameters ($\al$, $\niso$, and the normalization), which corresponds to
$\Ndof=38$ degrees of freedom.}
\label{tab:chimin}
\end{table}  

%%%%%%%%%%%%%%%%%%%%%%%%%%%%%%%%%%%%%%%%%%%%%%%%%%%%%%%%%%%%%%%%%%%%%%%%%%

\subsection{Combining CMB and LSS constraints}\label{CMB+LSS}

In order to further constrain the parameter space of allowed models, we
will consider in this Section the constraints coming from large
scale structure observations. In particular, we will constrain the shape
of the power spectrum, by comparing to results from the analysis of
galaxy clustering, and its amplitude by resorting to constraints from
the local abundance of rich galaxy clusters.

As for the shape of the galaxy power spectrum, different determinations
have been realized in the last few years, both for projected
\cite{pow2d} and redshift \cite{pow3d} samples. Such analyses converge
to indicate that the observed galaxy power spectrum is well reproduced
by an adiabatic CDM--like $P(k)$, in a flat universe, with shape
parameter $0.2\lesssim \Gamma \lesssim 0.3$, for a scale invariant
primordial spectrum \cite{shape}.

According to the results of the previous
section on the CMB constraints, the relative contribution of the
isocurvature component of the fluctuations is always rather small. 
As a result, the shape of the purely adiabatic spectrum is never
significantly changed by the isocurvature component.
For this reason, in order to implement the constraint from the shape
of the galaxy power spectrum, we will simply require in the following
that the adiabatic component of our mixed fluctuation spectrum has a
shape parameter $\Gamma$ lying in the 0.2--0.3 range.

As for the amplitude of the power spectrum, a powerful constraint is
represented by the number density of nearby galaxy clusters. 
Since rich galaxy clusters involve a typical mass of the order
of $10^{15}h^{-1}M_\odot$, their number density is connected to the
amplitude of density fluctuations at scales of about
$10\,h^{-1}$Mpc. Analytical approaches, based on the method originally
devised by Press \& Schechter \cite{prsc}, show that the cluster
abundance actually constrains the quantity
$\tilde\sigma_8=\sigma_8\Omega_m^\beta$, where $\sigma_8$ is the
r.m.s. fluctuation amplitude within a top--hat sphere of $8\,h^{-1}$Mpc
and $\beta\simeq 0.4$--0.5, almost independent of the shape of the power
spectrum and of the presence of a cosmological constant term
\cite{sigom}.

Different analyses, based on the distributions of $X$--ray cluster
luminosities, $X$--ray temperatures and velocity dispersions of member
galaxies, converge to values of $\tilde\sigma_8$ in the range 0.5--0.6
\cite{clabu}. For definiteness, in the following we will use for our
analysis the expression
\begin{equation}
\sigma_8\,=\,(0.55\pm 0.05)\,\Omega_m^{-0.43+0.09\Omega_m}
\label{eq:sigom}
\end{equation}
where the errors are intended to formally correspond to a 90\%
confidence level, while the $\Omega_m$ dependence is that provided by
Girardi et al. \cite{Geal98} for a flat Universe, with
$\Omega_\Lambda=1-\Omega_m$.

\begin{figure*}
\centering
\vspace{-1.truecm}
\hspace*{-2.truecm}
\leavevmode\epsfysize=13cm \epsfbox{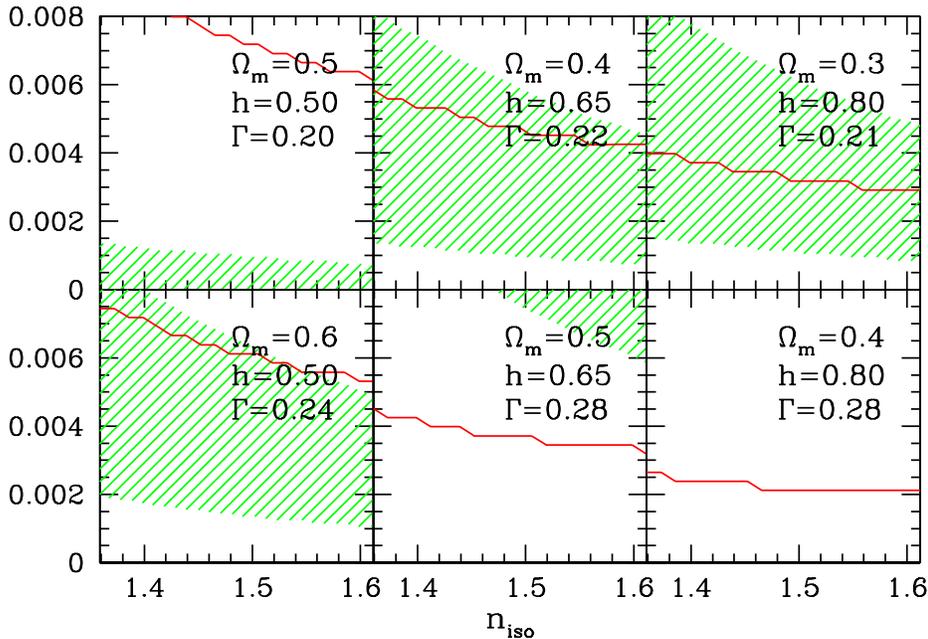}\\[3mm]
\vspace{-3.5truecm}
\caption[fig5]{\label{fig:lss} Constraints on the $\niso - \al$
plane from cluster abundance and CMB anisotropies. Each panel
corresponds to a choice for the $(\Omega_m,h)$ parameters which
satisfies the constraints on the shape parameter $\Gamma$. The
corresponding $\Gamma$ values for each model are also reported. The
shaded area correspond to the 90\% c.l. from the cluster abundance,
obtained according to eq.(\ref{eq:sigom}). The solid curves indicate
the 90\% c.l. upper limit on $\al$ from the CMB constraints (see
text); oscillations in the shape of these curves are due to
limitations in the numerical precision. }
\end{figure*}

We show in Figure \ref{fig:lss} constraints on the $\niso - \al$
parameter space by combining results from LSS and CMB data. Each panel
corresponds to a choice for the $(\Omega_m,h)$ parameters, among those
reported in Table \ref{tab:chimin}, which agrees with the measured
shape of the galaxy power spectrum.

As for the constraints from the abundance of local clusters, the 90\%
confidence level region [see Eq. (\ref{eq:sigom})] is shown by the
shaded area. We note that, as $\niso$ increases, the contribution of
the isocurvature fluctuations at the cluster scale also increases,
thus requiring a smaller $\alpha$ value to keep the power spectrum
amplitude at the level required by the cluster number density.

%%%%%%%%%%%%%%%%%%%%%%%%%%%%%%%%%%%%%%%%%%%%%%%%%%%%%%%%%%%%%%%%%%%%%%
% FIGURE 7: LSS
%%%%%%%%%%%%%%%%%%%%%%%%%%%%%%%%%%%%%%%%%%%%%%%%%%%%%%%%%%%%%%%%%%%%
%%%%%%%%%%%%%%%%%%%%%%%%%%%%%%%%%%%%%%%%%%%%%%%%%%%%%%%%%%%%%%%%%%%%%%

In order to establish confidence levels for model exclusion from the
analysis of CMB anisotropies we consider the quantity $\Delta
\chi^2=\chi^2-\Xmin^2$, where $\Xmin^2$ is the minimum value
as reported in Table \ref{tab:chimin}. We assume the $\chi^2$
statistics for $\Ndof=38$ to be normally distributed with mean
$\Ndof$ and r.m.s. scatter given by $\sqrt{2\Ndof}$. Accordingly,
all the models displayed in Figure \ref{fig:lss} correspond to an
acceptable value of $\Xmin^2$. The solid curve indicates the 90\%
c.l. upper limit on $\al$, which corresponds to $\Delta\chi^2=4.61$
for two significant fitting parameters.

As a general result, it is interesting to note that large scale
structure constraints significantly contribute to further restrict the
range of the allowed parameter space. For instance, the two models with
$(\Omega_m,h)=(0.5,0.65)$ and $(0.4,0.80)$ are now ruled out, since they
can not satisfy at the same time both the CMB and local cluster
abundance constraints, while the model with $(\Omega_m,h)=(0.5,0.50)$ is
constrained by the cluster abundance to have $\alpha\lesssim 0.0015$.

It is worth reminding here that the constraint of Eq. (\ref{eq:sigom})
from the local cluster abundance has been derived in the literature
under the assumption of Gaussian statistics for the density
fluctuations. On the other hand, the density fluctuations predicted by
our model are given by a scale dependent superposition of a Gaussian
adiabatic components and of a non--Gaussian isocurvature component,
whose probability density function (PDF) corresponds to a $\chi^2$ model
with one degree of freedom. Although the extension of the
Press--Schechter formalism to non--Gaussian statistics has been pursued
by different authors \cite{nong}, such attempts concentrated on
scale--independent PDF models. In our case, we expect that the positive
skewness of the $\chi^2$--distribution should ease the formation of
galaxy clusters, for a fixed $\tilde\sigma_8$, as a consequence of the
broader high density tail in the PDF for the isocurvature
component. Therefore, the net effect would go in the direction of
decreasing the required fluctuation amplitude at the cluster scale and,
thus, to somewhat increase the allowed $\alpha$ values. In any case,
since the isocurvature component is always constrained to be relatively
small even at the cluster scales ($\alpha_{\rm clus}\lesssim 0.15$; cf.
Fig.~\ref{fig9}), we are confident that our assumption of Gaussian
statistics should be a sensible one.

\subsection{The future CMB experiments}

In this section we compute the estimates of the errors with which the
future satellite experiments MAP and Planck will determine the
isocurvature contribution to the CMB power spectrum.  The aim here is to
verify whether the future CMB data alone will be able to constrain small
isocurvature contributions when other cosmological parameters are
constrained at the same time.

In order to provide such an estimate, we resort to the Fisher
information matrix approach \cite{TegTeHe97}.  When no polarization is
considered, the Fisher information matrix is defined as
\bea
F_{ij} &=& \sum_l {\partial C_l \over \partial \lambda_i} Cov^{-1} 
{\partial C_l \over \partial \lambda_j} \,,
\label{eq:fishmat} \\
Cov &=& {2 \over {(2l+1) f_{\rm sky}}} \left( C_l + 
w^{-1}e^{l^2\sigma_b^2}\right)^2 \,.
\label{eq:cova}
\eea
Here $f_{\rm sky}$ is the fraction of sky covered, $\sigma_b$ is the
Gaussian beamwidth ($\sigma_b = \tfwhm/\sqrt{8\ln 2}$, and $\tfwhm$ is
the full width at half maximum), ${\lambda_i}$ is the set of parameter whose 
errors have to be determined,  and $w$ contains the information on the
detector resolution and sensitivity:
\be
w \,=\, {\sigma^2_{\rm pixel} \Omega_{\rm pixel} \over T_0^2} \,,
\ee
where $\Omega_{\rm pixel} \simeq \tfwhm^2$. 

In the case that also polarization is considered, the expression for the
Fisher and covariance matrices become more complicated, and we refer
here to Ref. \cite{ZalSpSel97} for the explicit expression.  We note
that in the expression above a perfect foreground subtraction is
assumed, so that the estimates found should be considered somehow ideal.
Also notice that the expression reported here strictly holds for
Gaussian perturbations.  Therefore, its application to the
Linde--Mukhanov model should be taken with some caution, especially in
the very large multipole regime (i.e., small scales), where the
non--Gaussian isocurvature contribution may be non negligible.

We estimate the expected errors on $\al$ and $\niso$ with and
without polarization. In the calculation of the Fisher matrix, we
considered as free parameters the Hubble constant $h$, the baryon
abundance $\omB h^2$, the normalization $C_{10}$, the reionization
optical depth $\tau$, the isocurvature/adiabatic ratio $\al$, and the
isocurvature spectral index $\niso$.  Eqs.~(\ref{eq:cova}) and
(\ref{eq:fishmat}) are estimated by computing $C_l$ for a fiducial
model with $\omO = 0.3$ (fixed), $h = 0.65$, $\omB h^2 = 0.019$,
$C_{10}=C_{10_{\rm COBE}}$ and $\tau =0.05$.

The precision in the determination of $\al$ and $\niso$ is reported in
table \ref{tab:MAPPl} for different fiducial values, and different
experiments.  For both experiments, we combine different channels in
order to give the estimates on the errors.  More precisely, we consider
the 3 highest frequency channels for MAP (40, 60, 90 GHz), and the
5 central frequency channels from Planck (70, 100 GHz channels from
LFI and the 100, 143, 217 GHz channels from HFI).  The values of the
beamwidths and sensitivities for the two experiments are reported in
table \ref{tab:MAPPlpar}, according to Refs. \cite{MAP,Planck}.

As a result, we find that the future satellite experiments will be able
to constrain the spectral index $ \niso$ much better than the present
data. Clearly, $\niso$ is better determined if a higher
isocurvature component is allowed, and for a given $\al$ a higher
fiducial isocurvature anti-tilt ensures a smaller uncertainty in the
parameter determination.  Considering polarization is also useful in
this respect, and typically reduces the errors by almost a factor of 2.
If the isocurvature contribution is as low as indicated by our present
analysis, $\niso$ will be anyway determined by Planck with an error of
about 0.1.

As for the constraints on $\al$, table \ref{tab:MAPPl} shows that if the
isocurvature contribution is quite large ($\al \simeq 0.015$) both MAP
and Planck will be able to detect it; on the contrary, if $\al$ is as
low as the present data seem to suggest, only Planck (with polarization
information included) will be able to claim a detection, while MAP will
only put an upper limit at the $0.002-0.003$ level.  Note that the use
of polarization data certainly helps in reducing the error, although we
don't find the big improvement claimed by \cite{EnKu99}.  This may occur
because we keep the scalar to tensor ratio $R$ fixed in this model, and
therefore we are not affected by the degeneracy of this parameter with
$\al$.  Polarization is useful in breaking the degeneracy between the
reionization parameter $\tau$ and the isocurvature contribution $\al$
and helps in reducing the errors especially in the case of the Planck
experiment.

%%%%%%%%%%%%%%%%%%%%%%%%%%%%%%%%%%%%%%%%%%%%%%%%%%%%%%%%%%%%%%%%%%%%%%%%%
%		TABLE OF PARAMETERS FOR MAP AND PLANCK
%%%%%%%%%%%%%%%%%%%%%%%%%%%%%%%%%%%%%%%%%%%%%%%%%%%%%%%%%%%%%%%%%%%%%%%%
\begin{table}[h]
\centering
\begin{tabular}{ccccc}
Experiment & frequency (GHz) &  $\tfwhm$  & $\sigma_p$ ($\mu$K) &
$\sigma_p^{pol}$ ($\mu$K) \\
\hline \\[-2mm]
MAP & 40 &  $0.47^\circ$ & 35 & 49.3\\
MAP & 60 & $0.35^\circ$  & 35 &  49.3   \\
MAP & 90 & $0.21^\circ$  & 35 &  49.3 \\
Planck-LFI& 70 & $14^{\prime}$ & 9.8 & 13.8 \\ 
Planck-LFI& 100 &$10^{\prime}$ & 11.72 & 16.5\\
Planck-HFI& 100 &$10.7^{\prime}$ & 4.63 & -- \\
Planck-HFI& 143 &$8.0^{\prime}$ & 5.45 & 10.2\\
Planck-HFI& 217 & $5.5^{\prime}$ &11.7 & 26.2\\
\end{tabular}
\vspace*{2mm}
\caption{Beamwidths and sensitivities of the satellites experiments MAP 
and Planck.}
\label{tab:MAPPlpar}
\end{table}
%%%%%%%%%%%%%%%%%%%%%%%%%%%%%%%%%%%%%%%%%%%%%%%%%%%%%%%%%%%%%%%%%%%%%%%%

%%%%%%%%%%%%%%%%%%%%%%%%%%%%%%%%%%%%%%%%%%%%%%%%%%%%%%%%%%%%%%%%%%%%%%%%%%
%  			TABLE OF MAP & PLANCK DET. OF THE PARAMETERS
%%%%%%%%%%%%%%%%%%%%%%%%%%%%%%%%%%%%%%%%%%%%%%%%%%%%%%%%%%%%%%%%%%%%%%%%%

\begin{table}[h]
\begin{tabular}{cccccc}
\centering
fiducial & fiducial & MAP & MAP & Planck & Planck \\
$\al$ & $\niso$ & $\delta\al$ & $\delta\niso$  & $\delta\al$  & 
$\delta\niso$   \\
\hline  \\[-2mm]
%0.008 &   1.412 &  0.0050  & 0.115 & 0.0035  & 0.076 \\
%0.008 &   1.585 &  0.0027  & 0.053 & 0.0056  & 0.016 \\ 
%0.015 &   1.412 &  0.0061  & 0.071 & 0.0043  & 0.049 \\
%0.015 &   1.585 &  0.0035  & 0.036 & 0.0029 & 0.028 \\
% E: NEW ESTIMATES
0.002 &    1.439 &  0.0035  & 0.315 & 0.0024  &  0.209 \\
0.002 &    1.572 &  0.0022  & 0.195 & 0.0016  &  0.139 \\
0.008 &	   1.439 &  0.0044  & 0.094 & 0.0031  &  0.064 \\
0.008 &    1.572 &  0.0030  & 0.062 & 0.0023   & 0.047 \\
0.015 &    1.439 &  0.0054  & 0.059 & 0.0040   & 0.042 \\
0.015 &    1.572 &  0.0039  & 0.041 & 0.0031   & 0.031 \\
\hline
% HERE WITH POLARIZATION
0.002  &   1.439 & 0.0025  &  0.297 &  0.0013  &  0.126 \\
0.002  &   1.572 & 0.0015  &  0.172 &  0.0008  &  0.073 \\
0.008  &   1.439 & 0.0031  &  0.091 &  0.0017  &  0.039 \\
0.008  &   1.572 & 0.0020  &  0.056 &  0.0010  &  0.024 \\
0.015  &   1.439 & 0.0038  &  0.059 &  0.0020  &  0.025 \\
0.015  &   1.572 & 0.0026  &  0.038 &  0.0014  &  0.016 \\
\end{tabular}
\vspace*{2mm}
\caption{Estimation of the isocurvature mode parameters with the future
satellite experiments MAP and Planck. Top: polarization not
considered. Bottom: polarization considered.  The first two columns
indicate the fiducial value considered, and the others indicate the
estimated errors in the parameters.}
\label{tab:MAPPl}
\end{table}  

%%%%%%%%%%%%%%%%%%%%%%%%%%%%%%%%%%%%%%%%%%%%%%%%%%%%%%%%%%%%%%%%%%%%%%%%%%%%%
\section{Discussion}\label{discussion}

In this paper we have investigated the consequences of a CDM 
cosmology with mixed isocurvature and adiabatic initial conditions as
prescribed by a generic Linde--Mukhanov inflationary model.

We showed how the total spectra are modified by the isocurvature
contribution, as both the primordial spectral index for the isocurvature
power spectrum, $\niso$, and the isocurvature fraction, $\alpha$, are
varied within their allowed ranges.  In order to constrain the
parameters of the model we resorted to available data on the CMB
anisotropy, as well as on the large--scale structure constraints from
the shape of the galaxy power spectrum and the number density of nearby
galaxy clusters.  Observational constraints from CMB and LSS data have
been shown to provide complementary informations.  As a result, we found
that the allowed isocurvature contribution {\em at COBE scales} is
always very small, $\al\lesssim0.006$. Therefore, our results
generalizes to the $\niso>1$ case and strengthen the conclusions reached
in Ref. \cite{KaSuYa98} based on LSS constraints alone.

We note that, even allowing for the strong positive tilt of the
isocurvature component, the permitted isocurvature contribution at the
cluster scales is always small, with $\alpha_{\rm clus} \lesssim
0.15$. However, this contribution is expected to increase to 
$\sim 50\%$ level when we consider smaller scales, $\lesssim 1\hm$,
which are relevant for galaxy formation. Therefore, the resulting
$\chi^2$ (positive skewness) non--Gaussian statistics contributed by
the isocurvature fluctuations can play a significant role to ease the
galaxy formation at high redshift. The phenomenological implications
on galaxy formation of a small isocurvature contribution at the
COBE scales remains to be investigated in detail.

As for the determination of the isocurvature spectral index $\niso$, the
best fit to the CMB data is always provided by the lowest $\niso$
considered; although the limits from LSS and CMB data in
Fig. \ref{fig:lss} show a mild dependence on $\niso$. In this respect,
future CMB experiments can help to set more accurate limits on $\niso$,
with an optimistic estimate of $3\%$ and $1\%$ (high $\niso$ and $\al$)
and a more realistic one of $20\%$ and $8\%$ for MAP and Planck
respectively.

Finally, we showed that only the Planck experiment will be sensitive
enough to detect a possible non vanishing value for the $\alpha$
parameter within the limits already set by present CMB and LSS data.

\section*{Acknowledgements}
J.G.B. is supported by the Royal Society of London.  E.P. is a CITA
national fellow.  The authors wish to acknowledge SISSA for hosting all
of them during different phases of this work. The authors thank Andrei
Linde, Andrew Liddle, and Douglas Scott for generous discussions.

\end{document}